\documentclass[12pt]{article}
\pdfoutput=1
\usepackage{bm}
\usepackage{mathrsfs}
\usepackage{amsbsy}
\usepackage{graphicx}
\usepackage{subfigure}
\usepackage{amsmath}

\newcommand{\mpref}[1]{Figure.\ref{#1}}
\newcommand{\be}{\begin{equation}}
\newcommand{\ee}{\end{equation}}
\newcommand{\bea}{\begin{eqnarray}}
\newcommand{\eea}{\end{eqnarray}}
\begin{document}
	
	\begin{center}
		{\bf Page Curves and Islands in Charged Dilaton Black Holes}\\
		
		\vspace{1.6cm}
		
		{{\bf Ming-Hui Yu}$^{1}$,~ \bf Xian-Hui Ge$^{1,2}$}\let\thefootnote\relax\footnotetext{* Corresponding author. gexh@shu.edu.cn}\\
		\vspace{0.8cm}
		
		$^1${\it Department of Physics, Shanghai University, Shanghai 200444,  China} \\
		$^2${\it Center for Gravitation and Cosmology, College of Physical Science and Technology,
			Yangzhou University, Yangzhou 225009, China}
		\vspace{1.6cm}
		
		\begin{abstract}
			We study the Page curve in the four dimensional asymptotically flat eternal Garfinkle-Horowitz-Strominger dilaton black holes by applying the ``quantum extremal surface" prescription. The results demonstrate that without island, the entanglement entropy of Hawking radiation is proportional to time and divergent at late times. While taking account of the emergence of the island extending slightly outside the event horizon, the entanglement entropy of Hawking radiation stops growing and eventually reaches a saturation value, which is twice of the Bekenstein-Hawking entropy and consistent with the finiteness of the von Neumann entropy of eternal black holes. Moreover, we discuss the impact of the parameter $n$ and charge $Q$ on the Page time. The influence of $n$ on it can neglected when the ratio of the charge $Q$ to the black hole mass $M$ is sufficiently small, but the charge $Q$ has a significant impact on Page time.  Importantly, at the extremal case, the Page time becomes divergent or vanishing, in which the semiclassical theory needs further investigation.
		\end{abstract}
	\end{center}
\newpage
\tableofcontents
\newpage

\section{Introduction}
\qquad The black hole information paradox \cite{Paradox} is the fundamental problem of most significant field--quantum mechanics and general relativity-in theoretical physics, and has long been regarded as the key to the study of quantum gravity theory. Recently, there has been a breakthrough in the calculation of the fine-grained entropy of radiation for the evaporating black holes \cite{island1,island2,island3,island4,island5}, which means that the issue of black hole information can  be solved preliminarily. However, this is only done in frame of semiclassical physics, which seems to prompt us for an understanding of the microscopic mechanism of black holes.\\
\indent The black hole information issue was first initiated by Hawking in 1975, because Hawking radiation is the usual thermal radiation \cite{HR}, and the information that falls into the black hole will disappear forever after the black hole has evaporated. However, this conclusion is contrary to the basic assumption of quantum mechanics - the unitarity principle. If we assume that a black hole evaporates from a pure quantum state, based on the principle of unitarity, then at the end of the evaporation process it must still be in a pure quantum state instead of a mixed state. The behavior of the entropy of Hawking radiation in this process is described by the Page curve \cite{PC1,PC2}. Therefore, whether the black hole information issue can be solved is converted into whether the Page curve of the evaporating black hole can be reproduced.\\
\indent However, for a long time afterwards, physicists have been inconclusive on whether the evaporation process of a black hole satisfies the unitarity principle, until the discovery of the theory of AdS/CFT duality \cite{AdSCFT}. This theory provides a strong mathematical proof that the system in anti-de Sitter space (AdS) corresponds to the boundary conformal field theory (CFT). That is to say, the evaporation process of a black hole can be described by the boundary CFT, and CFT satisfies the principle of unitarity in quantum mechanics. Therefore, for an evaporating black hole, the entropy of Hawking radiation should follow the Page curve.\\
\indent Until recent years, the task to reproduce the Page curve is still difficult. Unfortunately, based on quantum theory considerations, the event horizon will be transformed into a high energy area - ``firewall", which is also hostile to the classical general relativity \cite{Firewall}. As the black hole evaporates, the amount of radiation becomes more and more at late times. In the case of an eternal black hole, the amount of radiation approaches infinite and leads to infinite amount of entropy, which obviously violates the principle of unitarity. Unitarity limits the maximum entropy of a black hole to be the Bekenstein-Hawking entropy \cite{Entropybound}. Therefore, the eternal black hole information paradox can be solved by expecting whether its Page curve grows to reach a bounded value, i.e. the Bekenstein-Hawking entropy.\\
\indent At present, the calculation of the Page curve for the Hawking radiation process was initially implemented using the semiclassical theory of the two-dimensional black holes under the context of asymptotically AdS spacetime in the JT (Jackiw-Teitelboim) gravity \cite{island1,island3,2D1}. Compared to four or high-dimensional systems, the structure of two-dimensional systems can provide more tractable analyses. Therefore, most of the current research is conducted in the two-dimensional gravity system. These studies have indicated that islands will appears in the end stage of the black hole evaporation. Under different context of spacetime, the location of island would be outside or inside the horizon. The effect of islands reduces the entanglement entropy and maintaining the finiteness of Bekenstein-Hawking entropy \cite{2D2,2D3,2D4,2D5}. Since the complexity of the structure of high-dimensional systems, which lacks symmetry analyses. The corresponding Page curves are more difficult to calculate, so there are less work in this situation. However, related work demonstrates that islands can work for high-dimensional spacetime \cite{island2}. Now, there are already many interesting and meaningful works \cite{Sch,RN,Extremal,Charged,Dilaton,higher,Massive,evaporating,Defect,Bath,Flat,Long range,charged dilaton}, especially the Page curve and island in Schwarzschild and Reissner-Nordstr\"om black holes \cite{Sch,RN}.\\
\indent One of the key applications of AdS/CFT duality is the holographic calculation of the entanglement entropy in quantum field theory, which is the famous holographic entanglement entropy conjecture proposed by Ryu and Takayanagi (RT) \cite{RT}. This conjecture transforms the calculation of the entanglement entropy in quantum field theory to a minimal surface in AdS spacetime. In the subsequent research, after considering all quantum corrections \cite{QRT1,QRT2,QRT3}, the RT proposal was extended to the ``Quantum Extremal Surfaces" (QES) prescription \cite{QES}. The ``QES" prescription is used to calculate the entanglement entropy and the results show that islands emerged at late times in the black hole evaporation, which is the reason for the decreasion of the radiation entropy. The behaviuor of entanglement entropy reproduces the Page curve, and the island formula for the fine-grained entropy of Hawking radiation is further derived by \cite{islandformula1,islandformula2}:
\begin{equation}
S(R)= {\rm min} \Bigg\{ {\rm ext}\bigg[\frac{ {\rm Area} (\partial{I})}{4G_N} +S_{\rm {matter}} ({R \cup I})\bigg]\Bigg\}, \tag{1.1} \label{1.1}
\end{equation}
with $G_N$ represents Newton constant, $R$ and $I$ are regions for radiation and island respectively. $\partial I$ is the boundary of the island and $S_{\rm matter}$ stands for the entanglement entorpy in the quantum matter field. Equivalently, the island formula can be derived from the replica trick for gravitational theories and the replica wormholes as saddle points exists in the island formula dervied from the gravitational path integral \cite{island1,Replicatrick}. \\
\indent Now, let us explain in detail our prescription and rules that must be followed. As mentioned above, the entanglement entropy after quantum correction is considered in ``QES" prescription, which is called the generalized entropy. It receives contribution from the Bekenstein-Hawking entropy of the island and the von Neumann entropy that evaluated on the $I$ and $R$ regions in the quantum matter field, which is written as
\begin{equation}
S_{\rm gen}=\frac{{\rm Area}(\partial{I})}{4G_N} +S_{\rm matter}(R\cup I).  \tag{1.2}  \label{1.2}
\end{equation}
The fine-grained entropy of the Hawking radiation of system is found by evaluating generalized entropy value under all extremal points that corresponds to the locations of the island, and pick the minimal value. If there is no extremal point, which means that the island is absent. We will see in later sections that it is the important interaction between the two contributions that causes a phase transition of the entanglement entropy at Page time and is consistent with unitarity.\\
\indent Since the most promising approach realizing the unified theories is the string theory at present, which might be reduced to the Einstein-Maxwell dilaton gravity in the low-energy limit, it is meaningful to study the charged dilaton gravity models. In this paper, we resorted to the two-sided geometry for quantum systems coupled with gravity, which also plays an important role in the development of gravity although there no holographic correspondences. We will address the information paradox in the four-dimensional asymptotically flat Garfinkle-Horowitz-Strominger dilaton (GHS) black hole and construct the Page curve and demonstrate that the entanglement entropy of Hawking radiation will converge at late times for the presence of the island, which is satisfies the unitarity. According to these results, we also study the influence of the parameter $n$ and charge $Q$ on the Page time.\\
\indent The content of the paper is as follows. In section \ref{Charged dilaton black holes }, we briefly introduce the properties of charged dilaton black holes. In section \ref{Entanglement entropy without island is divergent}, we calculate the entanglement entropy in the construction in the absence of island and reveal the information paradox for GHS black holes. In section \ref{Island limits the increase in entropy}, we analyze the behavior of generalized entropy at the early and late times. When we consider the construction of a single island, the unitary Page curve will reproduced. In section \ref{Page time and scrambling time}, based on the previous results, we discussed the relationship between the Page time to the parameter $n$ and charge $Q$. The conclusion and discussion are given in the last section. Finally, we approximate the entanglement entropy formula of two constructions in Appenix \ref{without island} and \ref{with island} respectively.

\section{Charged dilaton black holes }\label{Charged dilaton black holes }
\qquad In this section, we study the basic properties of Garfinkle-Horowitz-Strominger dilaton black holes, which is  a family of low-energy string theory representing static, spherically symmetric charged black holes \cite{GHS1}. Different from the Reissner-Nordstr\"om (RN) black hole in the classical general relativity, they are labeled by asymptotic value of the scalar dilaton, which could produce some interesting results.\\
\indent The action is generalized to the following form
\begin{equation}
I=\int d^4x \sqrt{-g}\big[R-2g^{{\mu}{\nu}}\nabla_{\mu}\varphi\nabla_{\nu}\varphi+e^{-2\alpha\varphi}F_{{\mu}{\nu}}F^{{\mu}{\nu}}\big], \tag{2.1} \label{2.1}
\end{equation}
where $R$ determines the Ricci scalar curvature, $\varphi$ denotes the dilaton, $F$ is the Maxwell tensor and the exponent $\alpha$ represents the coupling between the  electromagnetic field and dilaton.\\
\indent The metric of a static, spherically symmetric charged dilaton black hole can be written as follow
\begin{equation}
ds^{2}=-f(r)dt^{2}+f^{-1}(r)dr^{2}+R^{2}(d\theta^{2}+\sin^{2}\theta d\phi^{2}), \tag{2.2} \label{2.2}
\end{equation}
with the function $f(r)$ is defined by
\begin{equation}
f(r)=\bigg(1-\frac{r_+}{r}\bigg)\bigg(1-\frac{r_-}{r}\bigg)^{n}, \tag{2.3} \label{2.3}
\end{equation}
and
\begin{equation}
R^2(r)=r^{2}\bigg(1-\frac{r_-}{r}\bigg)^{1-n}, \tag{2.4} \label{2.4}
\end{equation}
where $n$ is a parameter has been introduced for convenience.
\begin{equation}
 n=\frac{1-\alpha^2}{1+\alpha^2}, \tag{2.5} \label {2.5}
\end{equation}
with $\alpha \in (0,1]$, so $n$ monotonically varies in the interval $[0,1)$. In these equations, Planck units are used $\hbar =G=c=k=1$ hereafter. The constants $r_+$ and $r_-$ are outer event horizon and inner horizon (Cauchy horizon) respectively. They could be expressend respectively as follows
\begin{equation}
r_+=M \Bigg [1+\sqrt{1-\frac{2n}{1+n}\bigg(\frac{Q}{M} \bigg)^2} \ \Bigg], \qquad  r_-=\frac{M}{n}\Bigg[1-\sqrt{1-\frac{2n}{1+n}\bigg(\frac{Q}{M}   \bigg)^2} \ \Bigg],   \tag{2.6}  \label{2.6}
\end{equation}
where $M$ is the mass and $Q$ is the charge in terms of  $r_+$ and $r_-$
\begin{equation}
M=\frac{1}{2}r_++\frac{n}{2}r_-,  \qquad Q^2=\frac{(1+n)}{2}r_+r_-.   \tag{2.7}  \label{2.7}
\end{equation}
Notice that the two horizons $r_+$ and $r_-$ coincide at
\begin{equation}
\frac{Q}{M}=\sqrt{\frac{2}{1+n}}, \tag{2.8}   \label{2.8}
\end{equation}
which corresponds to the extremal charged dilation black hole. The surface gravity of the event horizon is
\begin{equation}
\kappa_+=\frac{1}{2r_+} \bigg(1-\frac{r_-}{r_+} \bigg)^n.   \tag{2.9}   \label{2.9}
\end{equation}
Suppose that the black hole is in thermal equilibrium with a thermal bath of any temperature\footnote{Here, the location of the bath is where gravity is weak, and we ignore this weak effect.}, and the Hawking temperature is
\begin{equation}
T_+=\frac{\kappa_+}{2\pi}=\frac{1}{4\pi r_+}\bigg( 1-\frac{r_-}{r_+}\bigg)^n.      \tag{2.10}  \label{2.10}
\end{equation}
The area of the event horizon is
\begin{equation}
A_+=4 \pi R^2(r_+)=4\pi r_+^2 \bigg(1-\frac{r_-}{r_+} \bigg)^{1-n}, \tag{2.11}  \label{2.11}
\end{equation}
and the entropy of the black hole read as
\begin{equation}
S_{\rm BH}=\frac{A_+}{4G_N}=\frac{\pi r_+^2}{G_N}\bigg(1-\frac{r_-}{r_+} \bigg)^{1-n}. \tag{2.12}  \label{2.12}
\end{equation}
\indent Interestingly, the area of the event horizon and the black hole entropy becomes zero at the extremal limit ($r_+ = r_-$). We also noticed that in the limit case $n\to 1$, the metric restores the Reissner-Nordstr\"om (RN) black hole, while $n=0$, the metric corresponds to the Gibbons-Maeda-Garfinkle-Horowitz-Strominger (GMGHS) black hole. When the charge $Q=0$, the Schwarzschild black hole is recovered. The singularity of GHS black hole $r=0$ corrseponds to a negative value of $r$ in the original Schwarzschild metric \cite{Ge}.

\section{Entanglement entropy without island is divergent}\label{Entanglement entropy without island is divergent}
\begin{figure}[htb]
\centering
\includegraphics[scale=0.30]{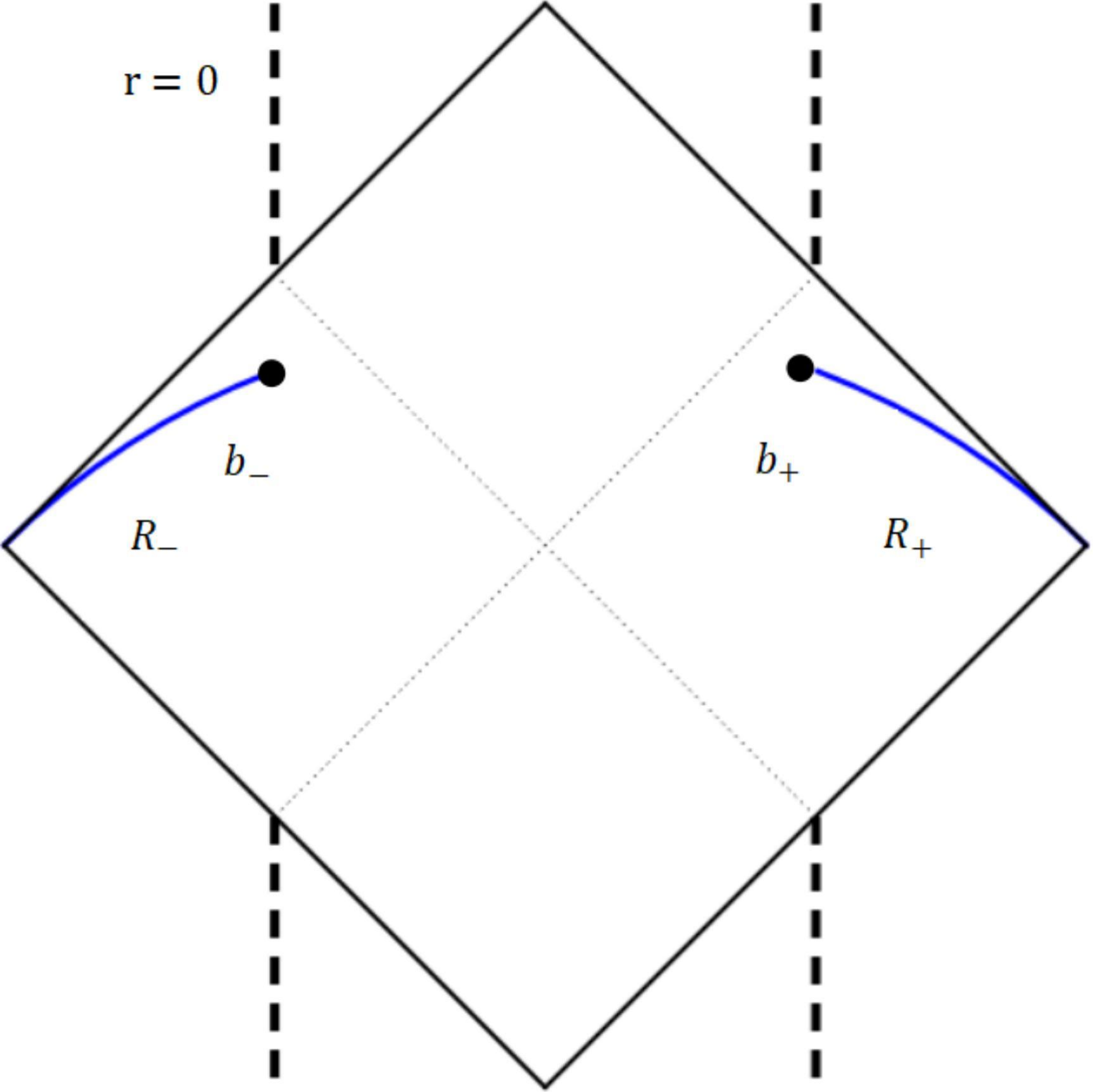}
\caption{\label{penrose1}Penrose diagram for GHS black hole without island. The Hawking radiation located in the left and right wedge, which are marked as $R_-$ and $R_+$. Their boundary surfaces are labeled by $b_{\pm}$. }
\end{figure}

\qquad In this section, we will calculate the entanglement entropy of Hawking radiation in the construction of without island and will see the GHS black hole information paradox more intuitively at the late times. Notice that in the configuration without island, the first term of the generalized entropy \eqref{1.2} is vanishing, and there is only contribution from radiation in the second term. Therefore, we have only two points that corresponds to boundaries of radiation regions at the left $R_-$ and the right $R_+$ (see \mpref{penrose1}).\\
\indent In four or high-dimensional spacetime, the entanglement entropy is usually unknown. However, the Hawking radiation of the system can be described by two-dimensional $s$-wave approximation for an observer very far away from it. Apart from that, we assume that the system in a pure quantum state at the initial time. In this case, the entanglement entropy of region $[b_-,b_+]$ is equal to entanglement entropy of the radiation, which can be written as (see Appendix \ref{without island})
\begin{equation}
S_R({\rm without}\ {\rm island})=S_{{\rm matter}}(R)=\frac{c}{3}\cdot \log[d(b_+,b_-)], \tag{3.1} \label{3.1}
\end{equation}
where $c$ is the centeral charge in CFT and $d(b_+,b_-)$ denotes the distance between the points $b_+$ and $b_-$ in GHS geometry. Here, we set the spacetime coordinates for them as $b_{\pm}=(t,r)=(\pm t_b,b)$. On the another hand, we find that it is useful to employ Kruskal transformation for the metric \eqref{2.2}. Define the tortoise coordinate
\begin{equation}
r^{\ast}= \int \frac{1}{f(r)}dr, \tag{3.2} \label{3.2}
\end{equation}
and define the Kruskal coordinates $U$ and $V$
\begin{equation}
U=-e^{-\kappa_{+}u}, \qquad
V=+e^{+\kappa_{+}v}, \tag{3.3} \label{3.3}
\end{equation}
where $\kappa_+$ is the surface gravity of the event horizen given in \eqref{2.9} and $u=t-r^{\ast}, v=t+r^{\ast}$ respectively. Under the Kruskal transformation, the metric  becomes the following form
\begin{equation}
ds^2=-g^2(r)dUdV+R^2(r)d\Omega^2,  \tag{3.4} \label{3.4}
\end{equation}
where $d\Omega^2=d\theta^{2}+\sin^{2}\theta d\phi^{2}$ and the function $g^2(r)$ is called the conformal factor, which can be written as follows
\begin{equation}
g^2(r)=\frac{f(r)}{\kappa_+^2e^{2\kappa_+r^{\ast}}}.  \tag{3.5}  \label{3.5}
\end{equation}
\indent Followed the conformal mapping, by using Eq. \eqref{3.1}, the matter sector of the radiation entropy in the GHS geometry contributes as
\begin{equation}
S_{\rm matter}(R)=\frac{c}{6}\cdot \log \Big \{  g(b_-)g(b_+)\big[U(b_-)-U(b_+)\big]\big[V(b_+)-V(b_-)\big] \Big \}, \tag{3.6}  \label{3.6}
\end{equation}
where
\begin{equation}
\sqrt{g(x)g(y)\big[U(y)-U(x)\big]\big[V(x)-V(y)\big]}=d(x,y), \tag{3.7}  \label{3.7}
\end{equation}
denotes the geodesic distance between points $x$ and $y$ in the GHS geometry. Substituting coodinates of points $b_\pm$, we can obtain\footnote{We should be cautiously swap the signs of $U$ and $V$ in Eq. \eqref{3.3} when considering points in the left region in the \mpref{penrose1}, since these points correspond to $r<0$. }
\begin{equation}
S_{\rm matter}(R)=\frac{c}{6}\Bigg[\frac{4f(b)}{\kappa_+^2}\cosh^2(\kappa_+t_b)\Bigg]=\frac{c}{6}\log \Bigg[16r_+^2\bigg(1-\frac{r_-}{r_+}\bigg)^{-2n}\bigg(1-\frac{r_+}{b}\bigg)\bigg(1-\frac{r_-}{b}\bigg)^{n} \cosh^2(\kappa_+t_b) \Bigg]. \tag{3.8}  \label{3.8}
\end{equation}
\indent At the late times, we assume that $t_b\gg b(>r_+)$, so we can employ the approximation
\begin{equation}
\cosh (\kappa_+t_b)\simeq \frac{1}{2}e^{\kappa_+t_b}. \tag{3.9}   \label{3.9}
\end{equation}
The above result can be recast as
\begin{align*}
S_{\rm matter}(R)&\sim \frac{c}{3} \log [\cosh(\kappa_+t_b)]  \\
                 & \simeq \frac{c}{3} \kappa_+ t_b.                 \tag{3.10} \label{3.10}
\end{align*}
Obviously, the entropy increases linearly with time and becomes infinite at the late times (see \mpref{pagecurve}). Therefore, in the absence of island, there is no Page curve and information does not escapes the black hole. The entanglement entropy will eventually be infinitely larger than the Bekenstein-Hawking entropy of the black hole, which is contradicts our previous assumptions. For an eternal black hole in the pure quantum state, its entanglement entropy is unchanged and with at most the Bekenstein-Hawking entropy. Therefore, the paradox appears. In next section, we will calculate the entanglement entropy of construction with an island and as long as is considered this construction, the unitary Page curve will be reproduced.

\section{Island limits the increase of entropy} \label{Island limits the increase in entropy}
\begin{figure}[htb]
\centering
\includegraphics[scale=0.30]{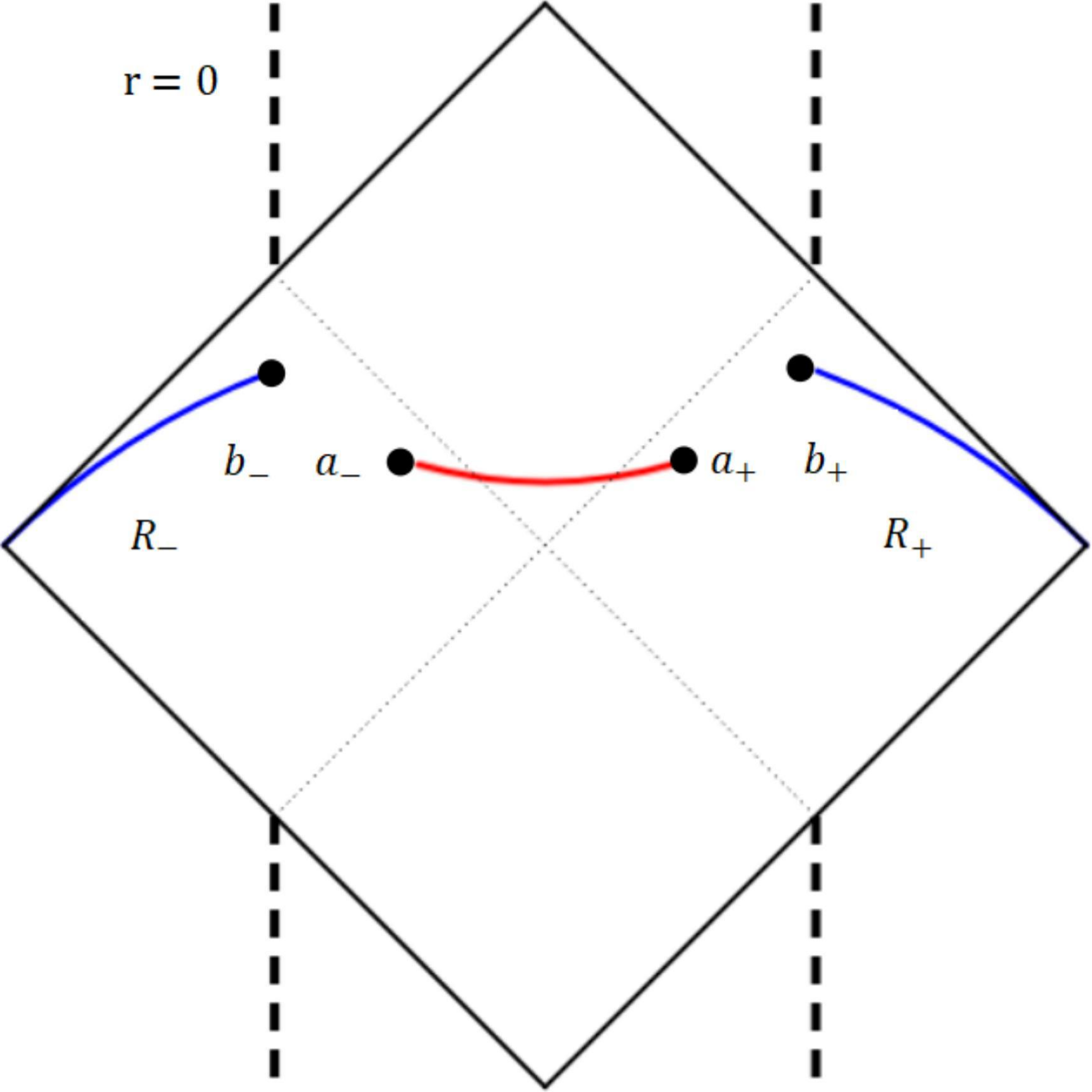}
\caption{\label{penrose2}Penrose diagram for the GHS black hole with a single island. The boundaries of the island are located at $a_-$ and $a_+$. The boundaries of the left and right radiation regions are indicated as points $b_-$ and $b_+$ respectively. Notice that the location of island where extends to outside of the event horizon.}
\end{figure}
\qquad In this section, we will calculate the entanglement entropy by considering the island. The construction is shown in \mpref{penrose2}\\
\indent We set the boundaries of the island located at $a_\pm=(t,r)=(\pm t_a,a)$, in this construction, the contribution from entropy of the matter field depends on the location of the radiation region $R$ relative to the event horizon. Besides, we only care about the behavior of entanglement entropy at late times, so the boundary of entanglement region $R$ is far away from the event horizon, i.e. $b\gg r_+$. We still assume that $s$-wave approximation to make sense. Under these approximations, the formula of the entanglement entropy that contributed from regions of radiation $R$ and island $I$ can be obtained as (see Appendix \ref{with island})
\begin{equation}
S_R({\rm with}\ {\rm island})=S_{{\rm matter}}(R\cup I)=\frac{c}{3}\cdot \log \Bigg[\frac{d(a_+,a_-)d(b_+,b_-)d(a_+,b_+)d(a_-,b_-)}{d(a_+,b_-)d(a_-,b_+)}\Bigg], \tag{4.1} \label{4.1}
\end{equation}
and the generalized entropy \eqref{1.2} in this construction is
\begin{equation}
S_{{\rm gen}}=\frac{2\pi R^2(a)}{G_N}+S_{{\rm matter}}(R\cup I). \tag{4.2} \label{4.2}
\end{equation}
Where $R^2(a)$ is defined in \eqref{2.4}, by using Eq. \eqref{3.7}, we can obtain the explicit expression of the generalized entropy
\begin{align*}
S_{\rm gen}=&\frac{2\pi R^2 (a)}{G_N}+\frac{c}{6}\log \Bigg \{ 2^8 r_+^4 \bigg( 1-\frac{r_-}{r_+} \bigg)^{-4n} \bigg(1-\frac{r_+}{a} \bigg) \bigg(1-\frac{r_+}{b} \bigg) \Bigg [ \bigg(1-\frac{r_-}{a}  \bigg) \bigg(1-\frac{r_-}{b}\bigg) \Bigg ]^n \\
            &\cosh^2(\kappa_+t_a) \cosh^2(\kappa_+t_b) \Bigg \}+\frac{c}{3}\log  \frac{\cosh [\kappa_+(r^\ast (a)-r^\ast (b))] -\cosh[\kappa_+(t_a-t_b)]}{\cosh  [\kappa_+(r^\ast (a)-r^\ast (b))]+\cosh[\kappa_+(t_a+t_b)]}, \tag{4.3}  \label{4.3}
\end{align*}
where $r^{\ast}(a,b)$ is the tortoise coordinate defined in Eq. \eqref{3.2}.\\
\indent The first term in Eq. \eqref{4.3} corresponds to the contribution from the area of the island. The second term corresponds to the entropy of the matter outside the radiation region $R$, and the last term is the entropy inside the island $R\cup I$. The fined-grained entropy of Hawking radiation is given by extremizing $S_{\rm gen}$ on top of all the extremal surfaces and choosing the minimal value. In the next subsection, we will find that the entropy of radiation are limited at late times.

\subsection{Island absent at early times}
\qquad Let us study the early and late time behaviors of the entanglement entropy. Since the Eq. \eqref{4.3} is too complicated, some appropriate approximations should be taken. At early times, we assume that $t_a,t_b\ll r_+$ and pick up the boundaries of radiation regions far away the event horizon: $b\gg r_+$. Therefore, the last term of $S_{\rm gen}$ \eqref{4.3} can be ignored, i.e.
\begin{equation}
\frac{c}{3}\log  \frac{\cosh [\kappa_+(r^\ast (a)-r^\ast (b))] -\cosh[\kappa_+(t_a-t_b)]}{\cosh  [\kappa_+(r^\ast (a)-r^\ast (b))]+\cosh[\kappa_+(t_a+t_b)]}  \to 0,  \tag{4.4}  \label{4.4}
\end{equation}
and we have
\begin{equation}
\bigg(1-\frac{r_+}{b}\bigg)\bigg(1-\frac{r_-}{b}\bigg)^n \simeq 1,  \qquad \cosh^2[{\kappa_+(t_a,t_b)}]\simeq 1. \tag{4.5}  \label{4.5}
\end{equation}
Therefore, $S_{\rm gen}$ is reduced to
\begin{align*}
S_{\rm gen}(\rm {early \ times})&\simeq \frac{2\pi  R^2(a)}{G_N}+ \frac{c}{6}\log \bigg [2^8r_+^4\bigg(1-\frac{r_-}{r_+}\bigg)^{-4n}\bigg(1-\frac{r_+}{a}\bigg)\bigg(1-\frac{r_-}{a}\bigg)^n \bigg ]\\
                                  &= \frac{2\pi  R^2(a)}{G_N}+\frac{c}{6} \log [f(a)]+\frac{2c}{3}\log \bigg(\frac{2}{\kappa_+}\bigg), \tag{4.6} \label{4.6}
\end{align*}
where the function $f(a)$ is the Eq. \eqref{2.3} for $a$ and $\kappa_+$ is the surface gravity defined in Eq. \eqref{2.9}. Extremizing this expression respect to $a$, i.e. $\frac{\partial S_{\rm gen}(\rm early \ \rm times)}{\partial a}=0$.\\
\indent This equation has no solution\footnote{Actually, the scale of the island is calculated to be smaller than Planck length and we discard all Plank scale physics. Besides we can not choose the extreme surface into the Cauchy horizon.}. The result indicates that there is no real extrem point, i.e. there is nonvanishing ``QES" that makes the generalized entropy reaching the extremal value. This also confirms our expecation that island is absent at early times. Therefore, the entanglement entropy is completely determined by contribution from radiation and increases linearly with time, which is consistent with Eq. \eqref{3.10}.

\subsection{Island emerged at late times}
\begin{figure}[htb]
\centering
\includegraphics[scale=0.9]{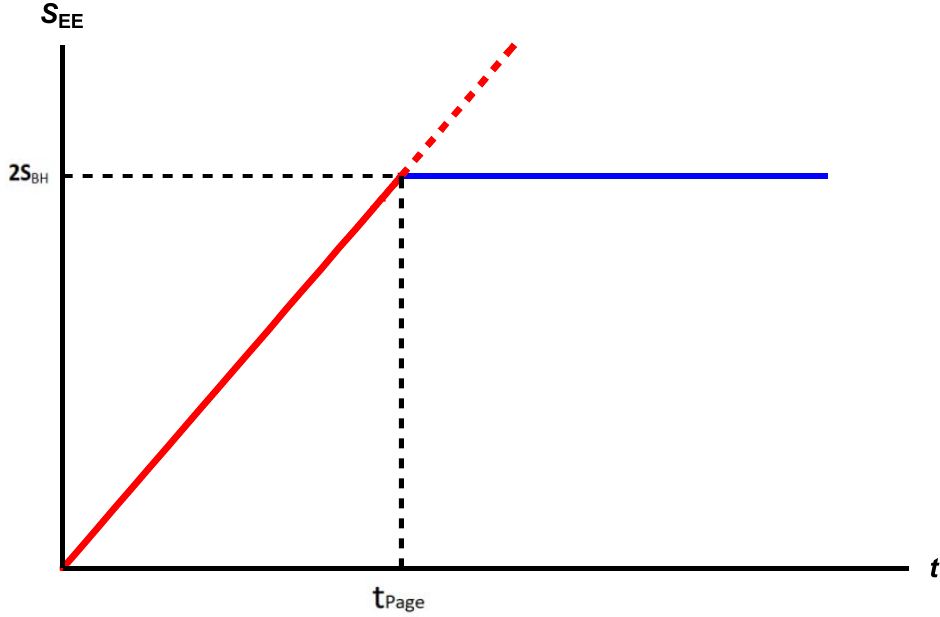}
\caption{The page curve for the eternal GHS black hole. The red dash line represents the entropy without island while the blue solid line stands the saturation vlaue of the entanglement entropy by considering the island.}
\label{pagecurve}
\end{figure}

\qquad Now, let us focus on the behavior of the entropy of Hawking radiation at late times. At the end stage of evaporation, because of the increasing amount of radiation, the entanglement entropy contributed by the matter part also grows. When the second term $S_{{\rm matter}}(R\cup I)$ in Eq.\eqref{1.2} grows up to ${\cal O}(G_N^{-1})$, the first term has the same order as it, then a phase transition will occur. The result is that the fine-grained entropy of Hawking radiation begins to decrease and the behavior of the entanglement entropy is constrained by unitarity as we expect.\\
\indent Consider the structure with an island, at late times, we assume that $t_a,t_b\gg b (>r_+)$. We first consider the time component of the Eq. \eqref{4.3}
\begin{equation}
S_{{\rm gen}}({\rm time})=\frac{c}{3}\log \Bigg \{ \cosh{(\kappa_+t_a)}\cosh{(\kappa_+t_b)} \frac{\cosh  [\kappa_+(r^\ast (a)-r^\ast (b))]-\cosh[\kappa_+ (t_a-t_b)]}{\cosh  [\kappa_+(r^\ast (a)-r^\ast (b))] +\cosh[\kappa_+ (t_a+t_b)]}\Bigg \}. \tag{4.7}  \label{4.7}
\end{equation}
Employed following approximations
\begin{equation}
\cosh{\kappa_+t_a} \simeq \frac{1}{2}e^{\kappa_+t_a},  \qquad
\cosh{\kappa_+t_b} \simeq \frac{1}{2}e^{\kappa_+t_b},  \tag{4.8} \label{4.8}
\end{equation}
and
\begin{equation}
\cosh{[\kappa_+(t_a+t_b)]}\gg \cosh{[\kappa_+(r^{\ast}(a)-r^{\ast}(b))}],  \tag{4.9}  \label{4.9}
\end{equation}
the expression above is reduced as
\begin{equation}
S_{{\rm gen}}({\rm time})\simeq \frac{c}{3} \log \Big \{ {\cosh  [\kappa_+(r^\ast (a)-r^\ast (b))}]-\cosh[\kappa_+ (t_a-t_b)] \Big \}. \tag{4.10}   \label{4.10}
\end{equation}
\indent We can easily find that when $\cosh[\kappa_+(t_a-t_b)]$ reach the minimum i.e. $t_a=t_b$, the expression obtains the maximal value. Set $t_a=t_b=t$ and substituting into the expression, we obtain the equation that no longer depends on time. This implies that the entanglement entropy converges at late times.\\
\indent Next, we calculate the explict form for the generlized entropy at late times. Invoking all of the approximations above \eqref{4.8} and \eqref{4.9}, we also approximate
\begin{equation}
\cosh [ \kappa_+ (r^\ast(a)-r^\ast(b))] \simeq \frac{1}{2}e^{\kappa_+[r^{\ast}(b)-r^{\ast}(a)]}.  \tag{4.11}  \label{4.11}
\end{equation}
Employed the above approximations, the generalized entropy in Eq. \eqref{4.3} can be reduced to
\begin{align*}
S_{{\rm gen}}&= \frac{2\pi  R^2(a)}{G_N}+\frac{c}{6}\log \bigg \{2^8r_+^4\bigg(1-\frac{r_-}{r_+}\bigg)^{-4n}\bigg(1-   \frac{r_+}{a}\bigg)\bigg(1-\frac{r_+}{b}\bigg)\Bigg [ (1-\frac{r_-}{a}\bigg)\bigg(1-\frac{r_-}{b}\bigg)\Bigg ]^{n}\\
&\cosh^2{\kappa_+t_a}\cosh^2{\kappa_+t_b} \bigg\}+\frac{c}{3}\log  \frac{\cosh [\kappa_+(r^\ast (a)-r^\ast (b))]-\cosh[\kappa_+ (t_a-t_b)]}{\cosh  [\kappa_+(r^\ast (a)-r^\ast (b))] +\cosh[\kappa_+ (t_a+t_b)]}\\
&\simeq  \frac{2\pi R^2(a)}{G_N}+\frac{c}{6}\log \bigg [ \frac{16f(a)f(b)}{\kappa_+^4}\cosh^4{\kappa_+t} \bigg]+\frac{c}{3}\log \frac{\frac{1}{2}e^{\kappa_+[r^\ast(b)-r^\ast(a)]}-1}{\frac{1}{2}e^{\kappa_+[r^\ast(b)-r^\ast(a)]}+\frac{1}{2}e^{2\kappa_+t}}\\
&\simeq \frac{2\pi  R^2(a)}{G_N}+\frac{c}{3}\log \Big[4g(a)g(b)e^{\kappa_+(r^\ast(a)+r^\ast(b))} \cosh^2 \kappa_+t \Big]+\frac{c}{3} \log \frac{\big[ e^{\kappa_+[r^\ast(b)-r^\ast(a)]}-2 \big]e^{-2\kappa_+t}}{e^{\kappa_+[r^\ast(b)-r^\ast(a)-2t]}+1}\\
&=\frac{2\pi  R^2(a)}{G_N}+\frac{c}{3} \log \frac{1-2e^{\kappa_+[r^\ast(a)-r^\ast(b)]}}{1+e^{\kappa_+[r^\ast(b)-r^\ast(a)-2t]}}+ \frac{2c}{3} \kappa_+r^\ast(b)+\frac{c}{3} \log[g(a)g(b)]\\
&=\frac{2\pi  R^2(a)}{G_N}+ \frac{2c}{3} \kappa_+r^\ast(b)+\frac{c}{3} \log[g(a)g(b)]+\frac{c}{3}\log \Bigg [  {1+\frac{-2e^{\kappa_+(r^\ast(a)-r^\ast(b))}-e^{\kappa_+[r^\ast(b)-r^\ast(a)-2t]}}{1+e^{\kappa_+[r^\ast(b)-r^\ast(a)-2t]}}}   \Bigg ]\\
&\simeq \frac{2\pi  R^2(a)}{G_N}+ \frac{2c}{3}\kappa_+r^\ast(b)+\frac{c}{3} \log[g(a)g(b)] - \frac{2c}{3}e^{\kappa_+[r^\ast(a)-r^\ast(b)]} -\frac{c}{3}e^{\kappa_+[r^\ast(b)-r^\ast(a)-2t] }.   \tag{4.12} \label{4.12}
\end{align*}
where in the first approximately equal sign, we used Eq. \eqref{4.11} and set $t_a=t_b=t$, and in the second approximately equal sign, we used the conformal factor defined in \eqref{3.5} to simplify. In the last line, we assumed the first order expansion in the logarithm: $\lim \limits_{x \to 0} \log (1+x) \simeq x$. \\
\indent Notice that the last term in the expression has dependence on the time $t$. However, it is the subleading order and quickly decays with time. Therefore, at late times, we can guarantee that the location of island is a constant and does not depend on time. Our current goal is to determine the location of the island. Extremizing this expression with respect to $a$ i.e. $\frac{\partial S_{{\rm gen}}}{\partial a}=0$, we obtain the location of the island as follows
\begin{equation}
a \simeq r_+ +{\cal O} \bigg(\frac{(cG_N)^2}{r_+^3}\bigg).  \tag{4.13}  \label{4.13}
\end{equation}
We can clearly see that the boundary of the island locates outside the event horizon, which is consistent with \mpref{penrose2}. Substituted this location into Eq.\eqref{4.3}, the real fine-grained entropy of Hawking radiation is obtained as
\begin{align*}
S_{{\rm EE}}&= \frac{2\pi r_+^2}{G_N}\bigg(1-\frac{r_-}{r_+}\bigg)^{1-n}+ \cdots\\
            &\simeq 2S_{\rm BH}.  \tag{4.14}    \label{4.14}
\end{align*}
The leading order of this expression is the Bekenstein-Hawking entropy of the black hole defined in \eqref{2.12}, which is produced by considering the construction of the island, subleading order and higher order terms are negligible compared to $S_{\rm BH}$.\\
\indent Based on these results, we briefly discuss the behavior of the generalized entropy. At early times, there is no island, the generalized entropy is dominated by the contribution from the matter part and grows with time. At late times, we consider the construction of the island that located outside the event horizon. The presence of the island is a necessary condition for generalized entropy to obtain the minimal value. The generalized entropy is saturated now and asymptotically becomes a constant, in which the leading order is the Bekenstein-Hawking entropy. Therefore, we reproduce the Page curve as shown in \mpref{pagecurve}.

\section{Page time and scrambling time} \label{Page time and scrambling time}
\qquad In this section, we will give the Page time and the scrambling time and discuss the influence of the parameter $n$ and charge $Q$ on the results. The Page time is defined as the moment when the radiation entropy of the entire system reaches maximum. For an eternal black hole, the entropy does not change after it. But for an evaporating black hole, its entropy will decrease after that. The relationship between its lifetime and the Page time was discussed in \cite{Lifetime}. The Page time of eternal GHS black hole is approximately the time when the entanglement entropy has maxmial value, which can be derived from the expression \eqref{3.10} and \eqref{4.14} of entropy for two constructions. Let them be approximately equal at late times and calculate as follows
\begin{equation}
t_{\rm Page}=\frac{6S_{\rm BH}}{c\kappa_+}=\frac{3S_{\rm BH}}{c\pi T_+}  \tag{5.1},  \label{5.1}
\end{equation}
where $\kappa_+$ and $T_+$ are the surface gravity and the Hawking temperature in Eq. \eqref{2.9} and Eq. \eqref{2.10} respectively. $S_{\rm BH}$ is the Bekenstein-Hawking entropy of the black hole given in Eq. \eqref{2.12}.\\
\indent The scrambling time is the recovery time of the information that fell into the black hole retrieved from the Hawking radiation \cite{Scramble1}. In the island prescription, the scrambling time corresponds to the time when information reaches the island. At time $t=0$, if we send a message from the radiation region $R (r=b)$, it will arrive at the island region $I (r=a)$ at time $t^{\prime}$. The scrambling time $t_{\rm scr}$ is defined as
\begin{equation}
t_{\rm scr}=t^{\prime}-t=r^{\ast}(b)-r^{\ast}(a)=\frac{1}{2\kappa_+} \log \Bigg [\frac{g^2(a)}{g^2(b)} \frac{f(b)}{f(a)} \Bigg] \tag{5.2}   \label{5.2}
\end{equation}
Substituting Eq. \eqref{4.13}, the expression can be written as
\begin{equation}
t_{\rm scr}\simeq \frac{1}{\kappa_+} \log \frac{r_+^2}{G_N}\simeq \frac{1}{2\pi T_+} \log S_{\rm BH}. \tag{5.3}  \label{5.3}
\end{equation}
Here, we set $r_+$ and $b$ has same order and the central charge $c$ is not very large: $c\ll \frac{1}{G_N}$. The result is consistent with \cite{Scramble2}, which is very small compared to the Page time. Therefore, we only focus on the Page time. \\
\indent The explicit expression of the Page time \eqref{5.1} through Eqs. \eqref{2.6} \eqref{2.10} and \eqref{2.12} is written as
\begin{align*}
t_{{\rm Page}}&=\frac{12\pi}{cG_N}r_+^3 \bigg(1-\frac{r_-}{r_+}\bigg)^{1-2n}\\
              &=\frac{12\pi}{cG_N}M^3 \Bigg[1+\sqrt{1-\frac{2n}{n+1} \bigg(\frac{Q}{M}\bigg)^2} \Bigg]^3 \Bigg ( 1-\frac{1-\sqrt{1-\frac{2n}{n+1}\big (\frac{Q}{M}^2\big)}}{n\Big[1+\sqrt{1-\frac{2n}{n+1}\big(\frac{Q}{M} \big)^2}\Big]}\Bigg )^{1-2n}.  \tag{5.4}   \label{5.4}
\end{align*}

\begin{figure}[htb]
\centering
\subfigure[\scriptsize{}]{\label{smallQ1}
\includegraphics[scale=0.5]{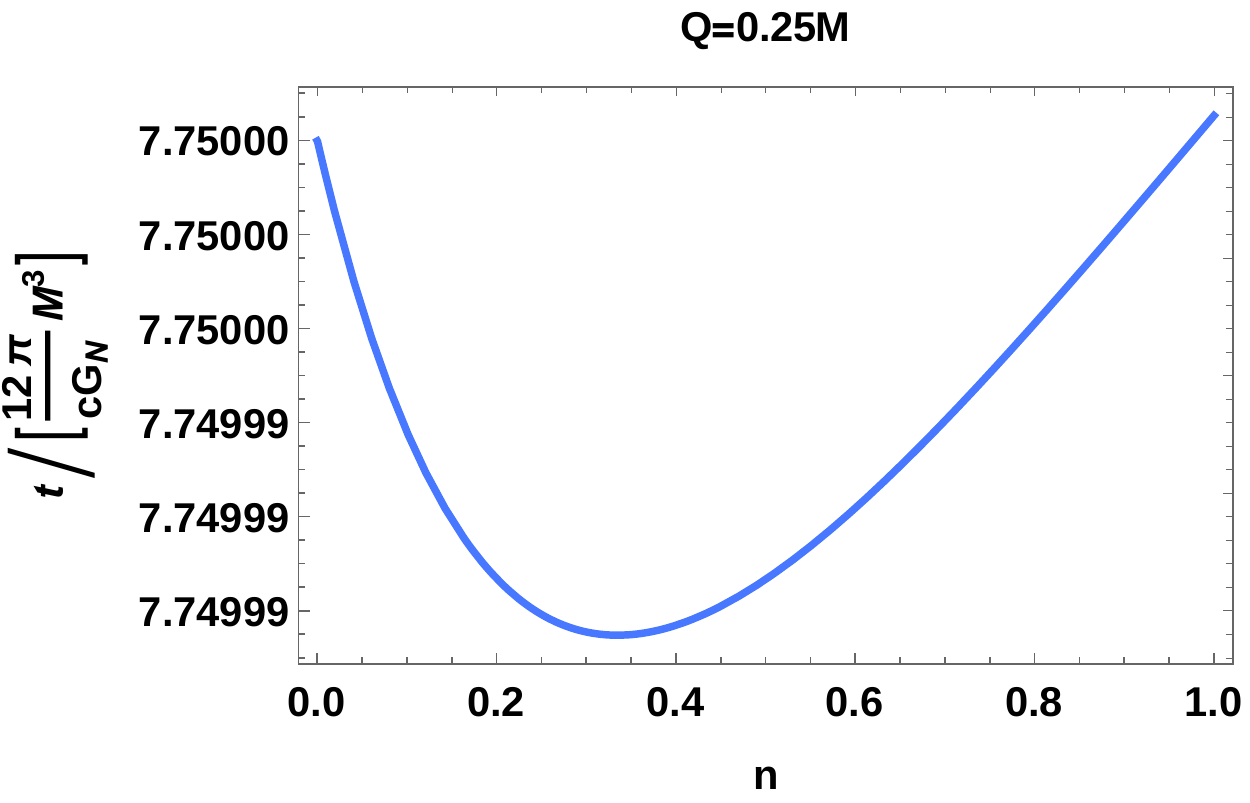}
}
\quad
\subfigure[\scriptsize{}]{\label{smallQ2}
\includegraphics[scale=0.5]{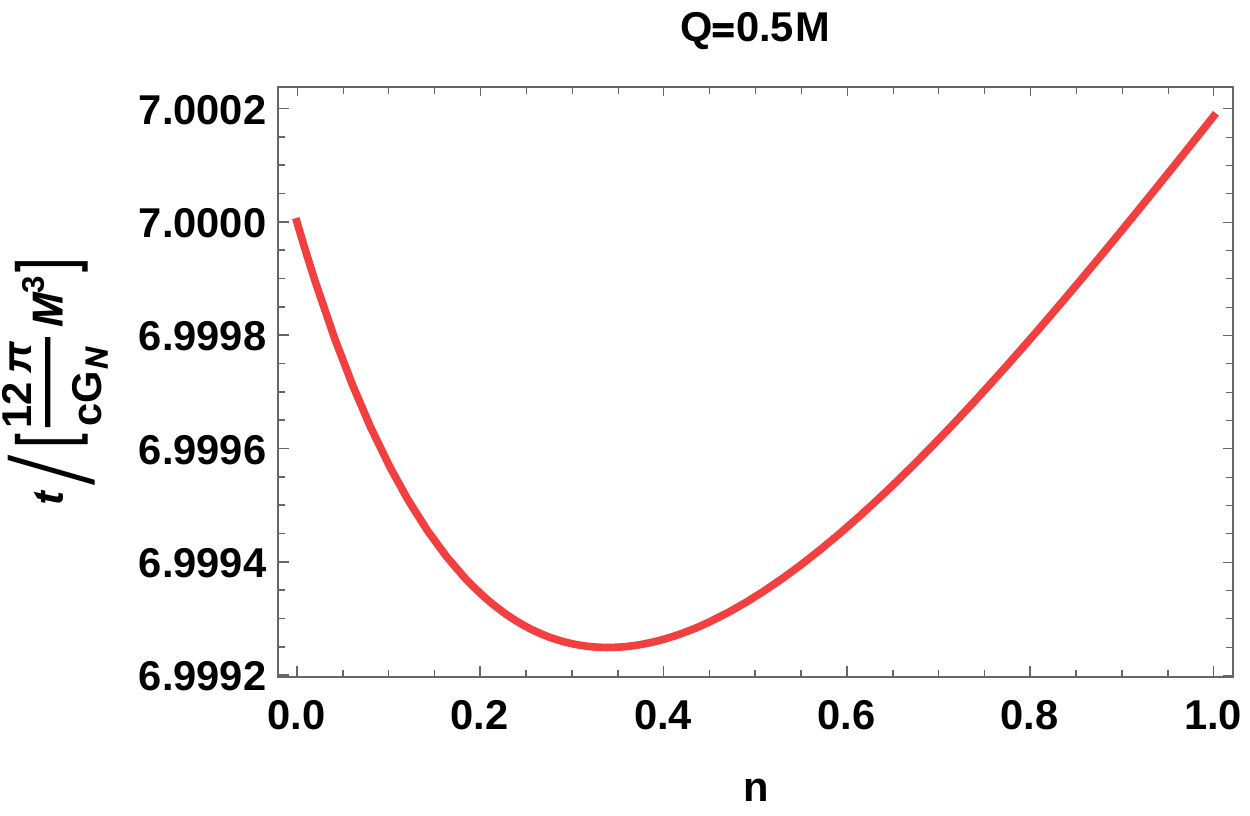}
}
\quad
\subfigure[\scriptsize{}]{\label{largeQ1}
\includegraphics[scale=0.5]{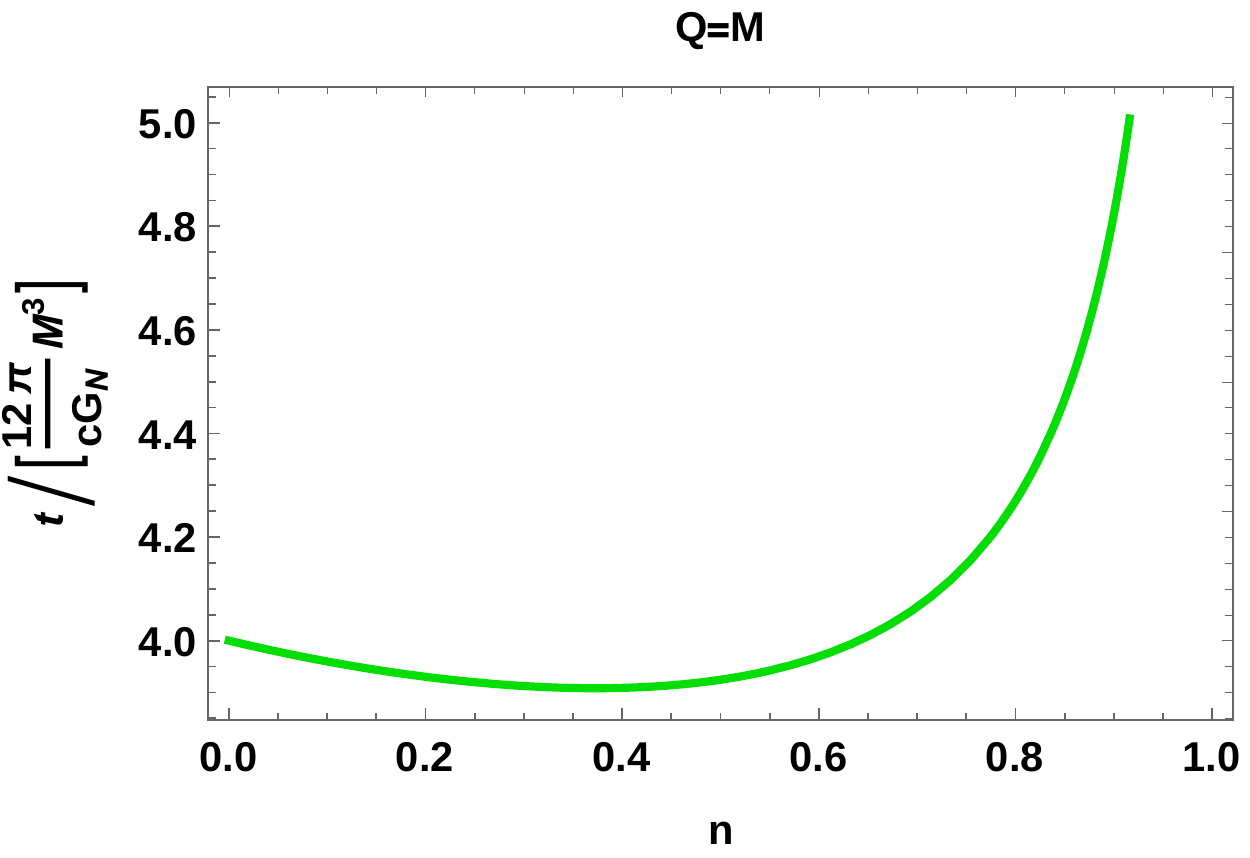}
}
\quad
\subfigure[\scriptsize{}]{\label{largeQ2}
\includegraphics[scale=0.5]{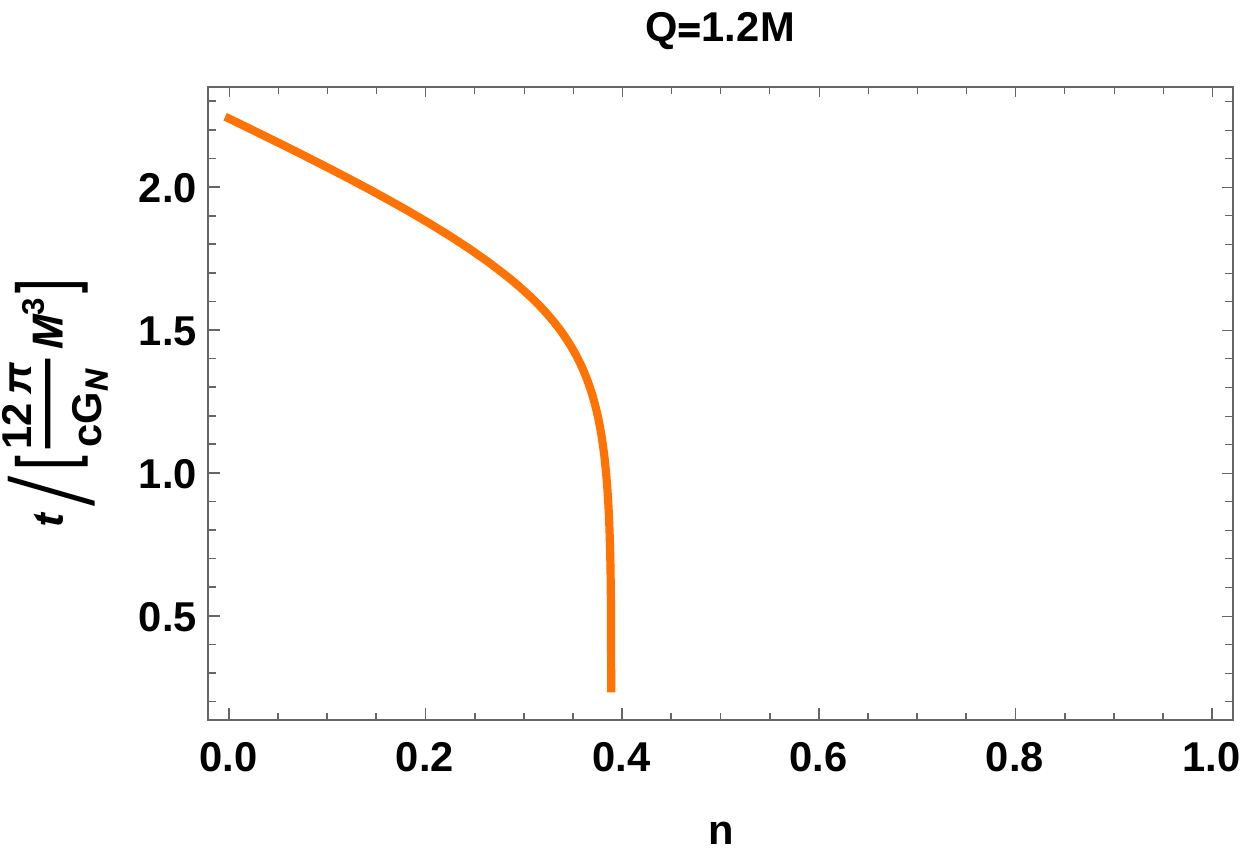}
}
\caption{The Page time by GHS black holes as a function for the parameter $n$ and four values of charge $Q$ : $Q=0.25M$ (a), $Q=0.5M$ (b), $Q=M$ (c), $Q=1.2M$ (d). In the third and last figure, the results becomes divergent or vanishing when $n$  approaches the critical value, which corresponds to extremal black holes. }
\label{FixCharge}
\end{figure}
\begin{figure}[htbp]
\centering
\subfigure[\scriptsize{}]{
\includegraphics[scale=0.5]{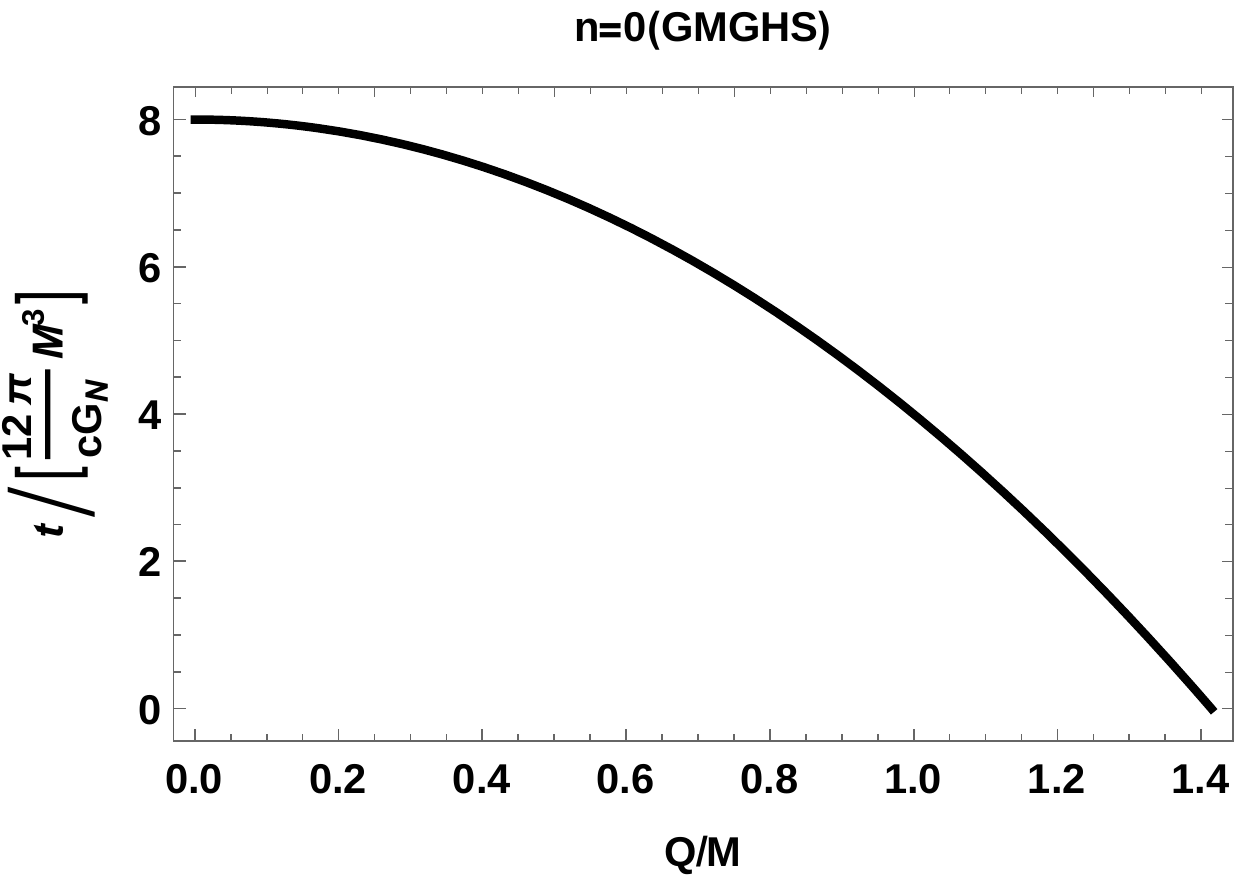}
}
\quad
\subfigure[\scriptsize{}]{
\includegraphics[scale=0.5]{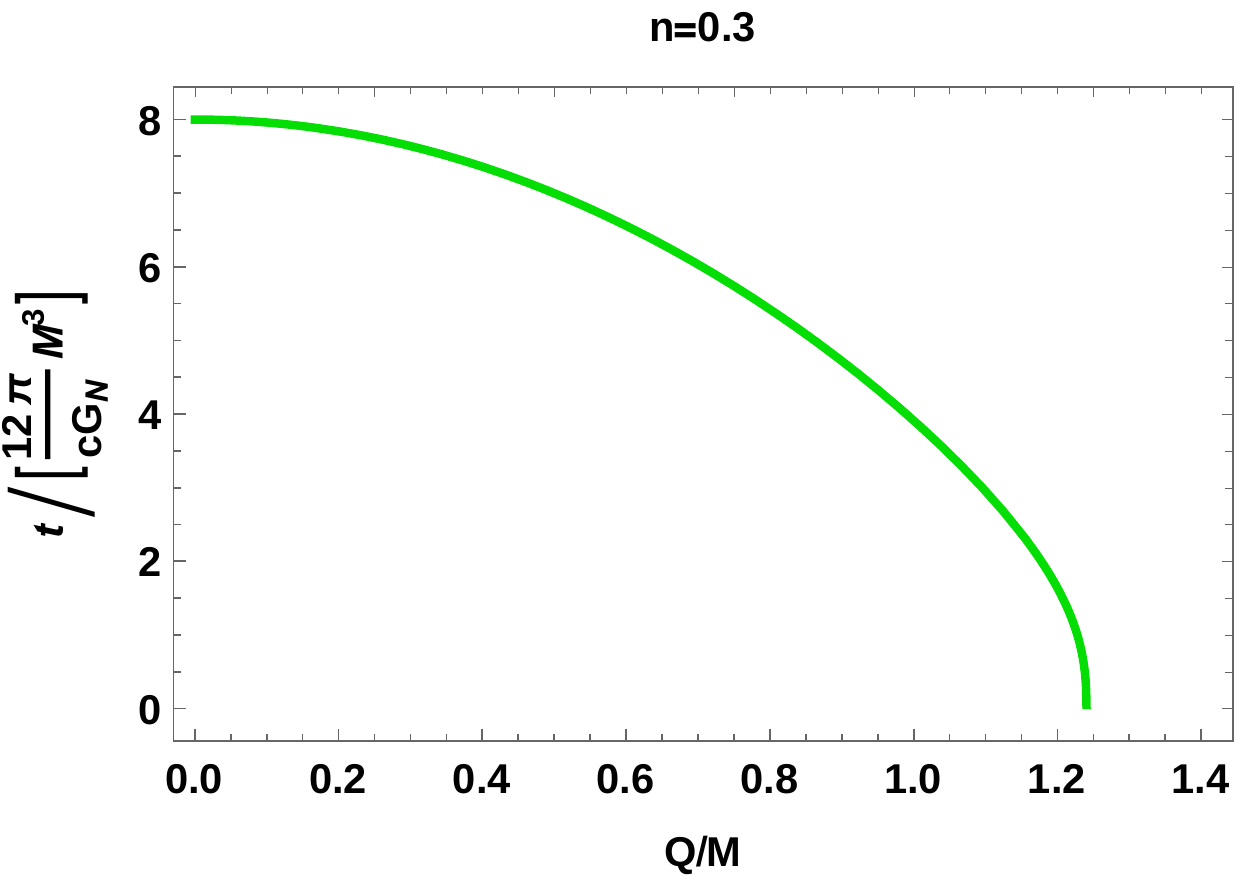}
}
\quad
\subfigure[\scriptsize{}]{
\includegraphics[scale=0.5]{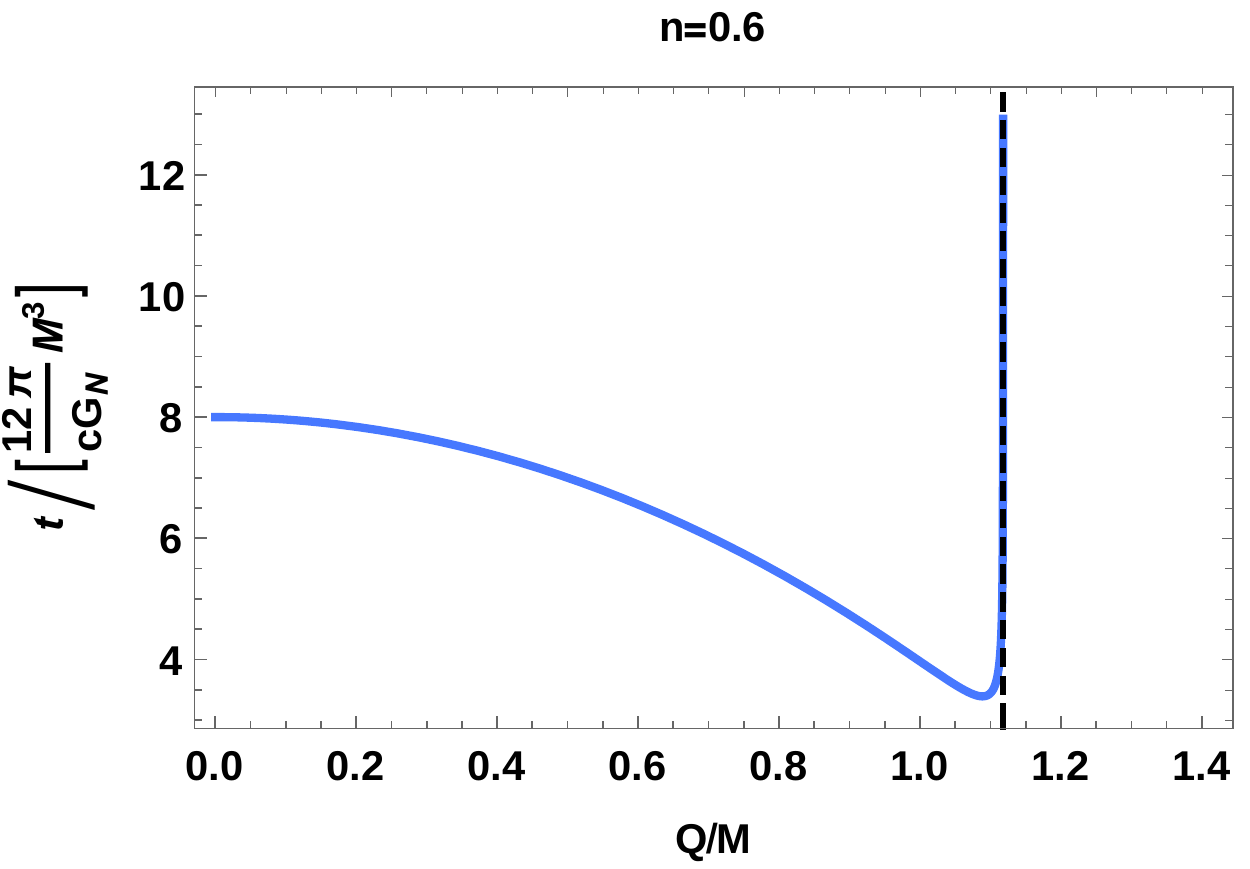}
}
\quad
\subfigure[\scriptsize{}]{
\includegraphics[scale=0.5]{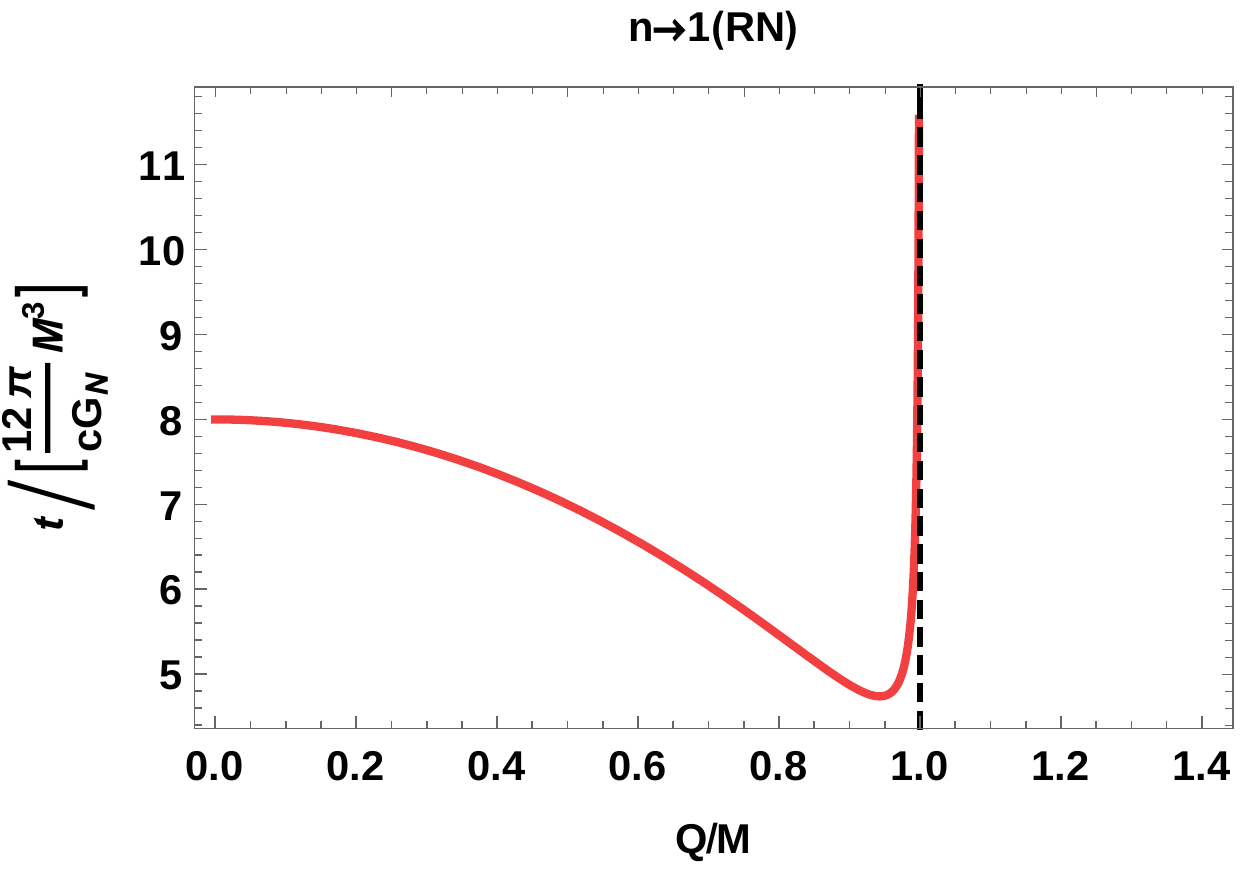}
}
\caption{The Page time of GHS black holes for the following values of $n$ : $n=0$ (GMGHS black holes) (a), $n=0.3$ (b), $n=0.6$ (c), $n\to1$ (RN black holes) (d). Onece again, we can find that the Page time is divergent or vanishing at large $Q$ in all figures.}
\label{Fixn}
\end{figure}
\begin{figure}[htb]
\centering
\subfigure[\scriptsize{}]{
\includegraphics[scale=0.45]{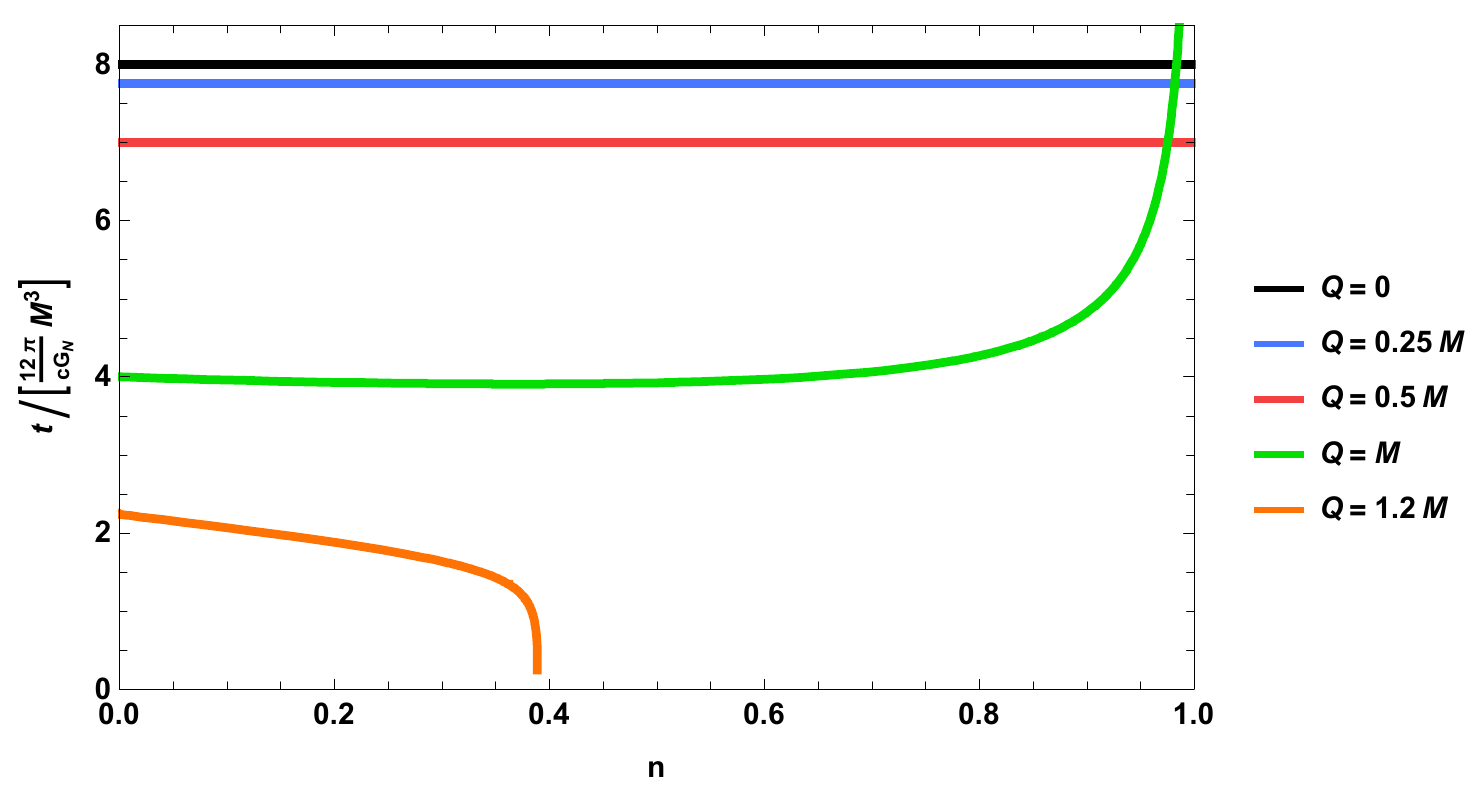}
}
\quad
\subfigure[\scriptsize{}]{
\includegraphics[scale=0.45]{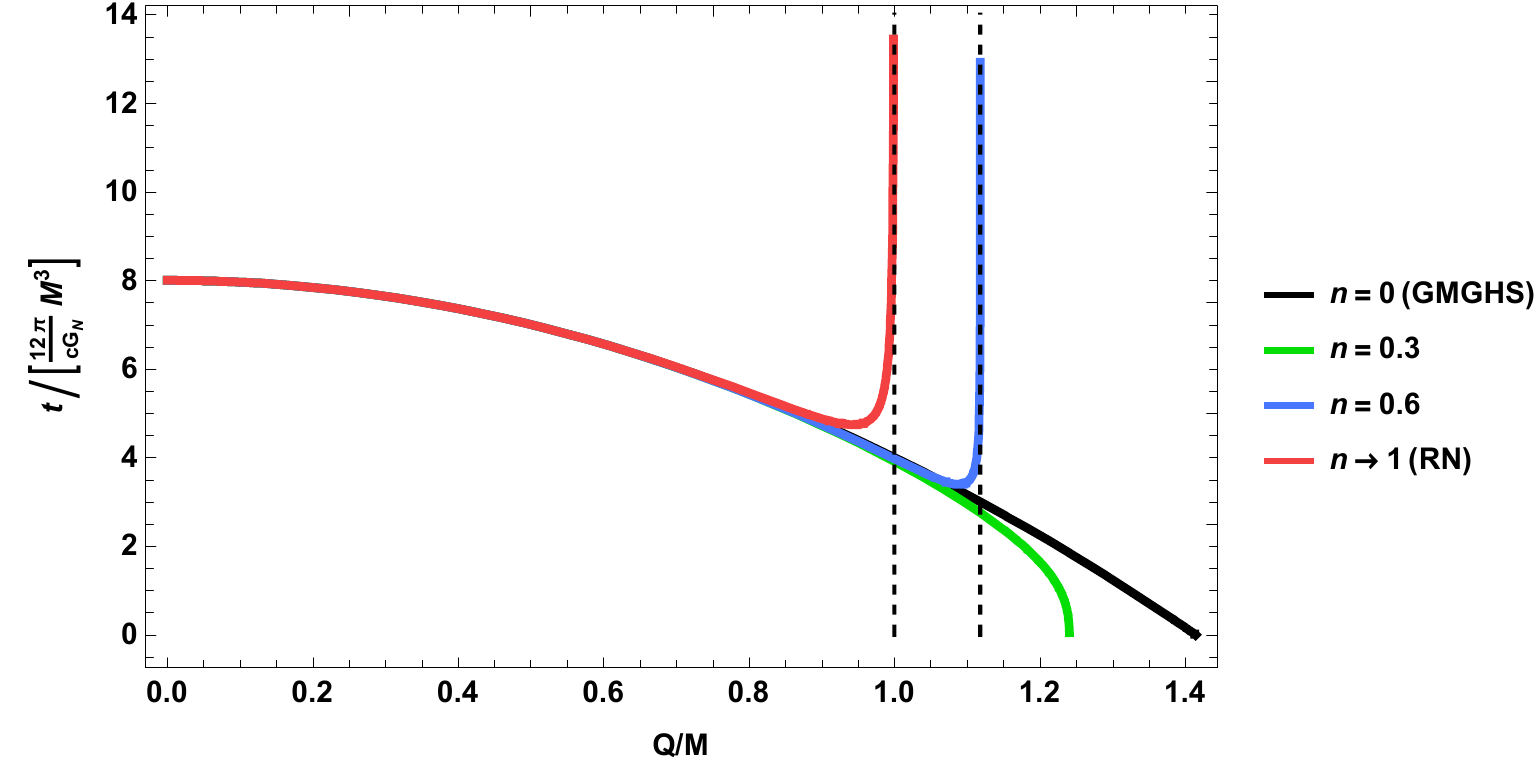}
}
\caption{The impact of the parameter $n$ and charge $Q$ on the Page time. On the left, the charge is fixed. The black line represtents the Page time of Schwarzschild black holes, which is a constant. On the right, the parameter is fixed. }
\label{All}
\end{figure}

\indent Firstly, fix the charge and investigate the influence of parameter $n$ on the Page time\footnote{From the Eq. \eqref{2.8}, the maximum charge carried by GHS black holes is $Q= \sqrt{2} M$.}. The results are shown in \mpref{FixCharge}, we can clearly know that when $Q/M$ is sufficient small, the interval of Page time varies with the parameter $n$ is very small (\mpref{FixCharge} \subref{smallQ1}, \subref{smallQ2}). Howerer, in the limit case $n= \frac{2M^2}{Q^2}-1$, the Page time is divergent or vanishing (\mpref{FixCharge} \subref{largeQ1}, \subref{largeQ2}), which corresponds to the extremal black hole. \\
\indent Subsequently, we discuss the effect of the amount of charge on the Page time of GHS black holes. The behavior of the Page time is plotted in \mpref{Fixn}. It decreases as charge $Q$ increases when $Q/M$ is sufficient small. But in the limit case $Q=\sqrt{\frac{2}{1+n}}M$ (i.e. the extremal GHS black hole), the Page time appears to divergent or zero, which is consistent with our previous discussion.\\
\indent \\
\indent The overall variation of the Page time with parameter $n$ and charge $Q$ is plotted in \mpref{All}. One the left, we can easily find only when the charge $Q$ is sufficient big, the parameter $n$ has a significant effect on the result,  in the particular case $Q=0$ (Schwarzschild black holes), Page time is a constant, which corresponds to \cite{Sch}. On the right, as the charge $Q$ increases, the Page time decreases, while at the extremal case $Q=\sqrt{\frac{2}{n+1}}M$, the Page time are shown as to be divergent or vanishing.\\
\indent To sum up, the impact of $n$ on Page time is neglected when the ratio of $Q/M$ is sufficiently small, but the charge $Q$ has a significant impact on it, especially at the extremal case, the Page time approaches to divergent or vanishing, which  implies that further investigate of the extremal case. We obtain data for all cases are recorded in Table.\ref{table}.
\begin{table}[htb]
\begin{center}
\linespread{2}\selectfont
\begin{tabular}{|c|c|c|}
\hline
Black Hole                     &  Requirement                                         &  The Page time $t_{\rm Page}$                                 \\[10pt] \hline
non-extremal GMGHS             &  $n=0,Q<\sqrt{2}M$                                   & $\frac{12\pi}{cG_N}r_+^3\big(1-\frac{r_-}{r_+} \big)$         \\[10pt] \hline
extremal     GMGHS             &  $n=0,Q=\sqrt{2}M$                                   & $\rm{vanishing}$                                              \\[10pt] \hline
non-extremal   GHS             &  $n\in(0,1),Q<\sqrt{\frac{2}{1+n}}M$               & $\frac{12\pi}{cG_N}r_+^3\big(1-\frac{r_-}{r_+} \big)^{1-2n}$  \\[12pt] \hline
extremal       GHS             &  $n\in(0,1),Q=\sqrt{\frac{2}{1+n}}M$               & $\rm{vanishing\ or \ divergent}$                              \\[10pt] \hline
non-extremal RN                &  $n\to1,Q< M$                                        & $\frac{12\pi}{cG_N}r_+^3\big(1-\frac{r_-}{r_+} \big)^{-1}$    \\[12pt] \hline
extremal RN                    &  $n\to1,Q=M$                                         & $\rm{divergent}$                                              \\[10pt] \hline
Schwarzschild                  &  $Q=0$                                               & $\frac{12\pi}{cG_N}r_+^3$                                     \\[10pt] \hline

\end{tabular}.
\caption{The results of the Page time with different condiations. The results of Schwarzschild black holes and RN black holes are consistent with \cite{Sch,RN,Extremal}. }
\label{table}
\end{center}
\end{table}

\section{Conclusion and Discussion}
\qquad In this paper, we study the black hole information issue in case of eternal black holes. For an eternal black hole, suppose that it is in the pure quantum state at the initial time, in the end of stage of evaporation, the amount of Hawking radiation tends to be infinite, which will produce infinite thermal radiation entropy. Therefore, this is beyond the Bekenstein entropy bound and conflicts with the unitarity of quantum mechanics. However, for a real evaporating black hole, the information issue is essentially the same as an eternal black hole. Considered the final stage of the evaporation process, the black hole completely evaporates and turns into the Hawking radiation. We assume that it has the same initial condition, the fine-grained entropy of radiation should drop to zero at the end, but the thermal radiation entropy is also much larger than the entropy bound. Therefore, in the case of eternal black holes, it is also significant to study the information paradox.\\
\indent In summary, based on a series of important solutions of the Einstein's equation in low-energy string theory, we studied the black hole information paradox in four-dimensional GHS spacetime. At early times, the system has not emitted enough Hawking radiation, the entanglement entropy of the system is contributed from matter sector and no island is formed. However, at late times, an island emergences outside the event horizon, which changed the construction of the system, the entanglement entropy is dominated by island region. The entire behavior of entanglement entropy can described from the Page curve in the \mpref{pagecurve}. Based on it, we calculated the Page time and the scrambling time and discussed the impact of parameter $n$ and charge $Q$ on the Page time (\mpref{All} and Table.\ref{table}). However, for the extremal black hole, the results are divergent or vanishing, which requires further investigation on the island prescription.\\
\indent We should also pay attention to the following issues. Initially, for the structure of the system, this paper only studies the situation of a single island. The construction of multiple islands may produce additional entanglement entropy, but their contribution is high-order and can be ignored. They probably appear at Page time, thus alleviating the phase transition of entanglement entropy \cite{Sch}. Besides, we assume that the eternal black hole to be in thermal equilibrium with a heat bath in asymptotically flat spacetime. However, the island ruler was first obtained from the black hole in AdS spacetime coupled to a weak gravitational bath. Therefore, one should be careful to investigate in the asymptotically flat spacetime. At last, from the aspect of information, the Page curve is obtained by considering the construction of the island, but how the information trapped on the island is connected to the outside of the radiation, a innovative conjecture is $\rm``ER=EPR"$ \cite{ER}. However, it still lacks a strong proof mathematically. Maybe we still need a complete theory of quantum gravity to explain the microscopic mechanism of black holes, which will reveal the mystery of black hole information.

\section*{Acknowledgement}
We would like to thank Yang Zhou, Chen-Yang Dong, Cheng-Yuan Lu and Yu-Qi Lei for helpful discussions. The study was partially supported by NSFC, China (grant No.11875184).\\

\section*{Note added}
After we have finished this paper, we receive a very interesting paper \cite{charged dilaton}, where the authors
introduce an alternative method on taking extremal limit of black holes. It would be
interesting to perform a similar analysis on our black hole background.

\section*{Appendix}
\begin{appendix}
\section{Entropy formula for the construction without island}\label{without island}
In this appendix, we give the entropy formula for the case without island. For the metric in two dimensions
\begin{equation}
ds^2=-e^{2\rho}dX^+dX^-, \tag{A.1}  \label{A.1}
\end{equation}
the matter entropy per unit area in two-dimensions is given by \cite{Without1,Without2}
\begin{equation}
S_{{\rm matter}}=\frac{c}{6}\cdot \log [d^2(x,y)],  \tag{A.2}  \label{A.2}
\end{equation}
where
\begin{equation}
d^2(x,y)=-[X^+(x)-X^+(y)][X^-(x)-X^-(y)]e^{\rho(x)}e^{\rho(y)}.  \tag{A.3}  \label{A.3}
\end{equation}
Supposing that the two-dimensional $s$-wave approximation is valid, we can use these equations to derive Eqs. \eqref{3.1} \eqref{3.6} and \eqref{3.7}.

\section{Entropy formula for the construction with an island}\label{with island}
In this appendix, we approximate the entropy formula for the case with an island. Due to the presence of the island, the system is divided into two disjoint intervals. The region of radiation is
\begin{equation}
R=[b_-,\infty_L)\cup[b_+,\infty_R), \tag{B.1}   \label{B.1}
\end{equation}
and the region of island is
\begin{equation}
I=[a_-,a_+].     \tag{B.2}   \label{B.2}
\end{equation}
In the theroy of free Dirac fermions, the explicit formula of non-universal entropy is obtained by \cite{With}
\begin{equation}
S_{{\rm fermions}}=\frac{c}{6}\cdot \log \Bigg[  \frac{\left|x_{21}x_{32}x_{43}x_{41}\right|^2}{\left|x_{31}x_{42}\right|^2 \Omega_1 \Omega_2 \Omega_3 \Omega_4}   \Bigg],   \tag{B.3}  \label{B.3}
\end{equation}
where $x_{ab}$ is the distance of two points $a,b$ and two disjoint interval is set to be
\begin{equation}
[x_1,x_2] \cup [x_3,x_4],    \tag{B.4}   \label{B.4}
\end{equation}
with $\Omega$ is the warp factor for the metric
\begin{equation}
ds^2=-\Omega^{-2}dX^+dX^-.     \tag{B.5}    \label{B.5}
\end{equation}
Therefore, employing the two-dimensional $s$-wave approximation, choosing $x_2,x_3$ as the left and right boundaries of the island region, and $x_1,x_4$ as the left and right boundaries of the radiation region, we can obtain Eq. \eqref{4.1}.
\end{appendix}

%\bibliographystyle{plainnat}
%\bibliography{ref}
\end{document}